\documentclass[prl,twocolumn,showpacs,superscriptaddress,amsfonts]{revtex4}

\usepackage{psfig}

\renewcommand{\vec}[1]{{\rm\bf #1}}
\newcommand{\tr}{\mathop{\rm tr}}
\newcommand{\cC}{\mathcal{C}}
\newcommand{\cD}{\mathcal{D}}

\begin{document}

\title{Dynamic localization in quantum dots: analytical theory}
\author{D.~M.~Basko}
\email{basko@ictp.trieste.it}
\affiliation{The Abdus Salam International Centre for Theoretical Physics,
Strada Costiera 11, 34100 Trieste, Italy}
\author{M.~A.~Skvortsov}
\affiliation{Landau Institute for Theoretical Physics,
2 Kosygina Street, 117940 Moscow, Russia}
\author{V.~E.~Kravtsov}
\affiliation{The Abdus Salam International Centre for Theoretical Physics,
Strada Costiera 11, 34100 Trieste, Italy}
\affiliation{Landau Institute for Theoretical Physics,
2 Kosygina Street, 117940 Moscow, Russia}

\date{\today}
\begin{abstract}
We analyze the response of a complex quantum-mechanical system  (e.g., a
quantum dot) to a time-dependent perturbation $\phi(t)$.
Assuming the dot energy spectrum and the perturbation to be described
by the Gaussian Orthogonal Ensemble of random matrices we find the
quantum corrections to the
energy absorption rate as a function of dephasing time $t_{\varphi}$.
If $\phi(t)$ is a sum of $d$ harmonics with {\it incommensurate} frequencies,
the quantum corrections behave similarly to
those of conductivity $\delta\sigma_{d}(t_{\varphi})$  for the
$d$-dimensional Anderson model of the orthogonal symmetry class.
For {\em periodic} perturbations,
the leading quantum corrections are generically absent
as in the systems of the unitary symmetry class. Exceptions are the harmonic
perturbation and all other periodic perturbations $\phi(t)$ that obey the
generalized time-reversal condition $\phi(-t+\tau)=\phi(t+\tau)$, where $\tau$
is a certain shift. Such cases fall into the quasi-1d orthogonal universality class.
\end{abstract}


\maketitle

The process of energy absorption by a quantum system with a
time-dependent Hamiltonian underlies a large part of
modern physics, both fundamental and applied.
A generic Hamiltonian can be written in the form:
\begin{equation}
\label{Ht=}
  \hat{H}(t)=\hat{H}_{0}+\hat{V}\phi(t) ,
\end{equation}
where we explicitly separated the time-independent part~$\hat{H}_{0}$
and the external perturbation~$\hat{V}$ with the time dependence specified
by a given function~$\phi(t)$. Most often the relevant
case is that of the classical ohmic Joule absorption. The simplest nonlinear
effects that restrict the absorption rate are the saturation effects that
originate from
an upper bound on the spectrum of~$\hat{H}_0$ (as in the textbook
example of a two-level system).

In the past two decades attention of the scientific community was
drawn to a different and much less trivial example of saturation when
the spectrum of $H_{0}$ is essentially unlimited in the energy space, yet,
after a certain time the absorption stops.
This so-called dynamic localization~(DL) in the energy space was
observed in numerical simulations on the kicked quantum rotor (KQR) --
particle on a circle with $\hat{H}_0=-\partial^2/\partial\theta^2$ and
$\phi(t)$~being a periodic sequence of $\delta$-pulses~\cite{Casati79},
as well as in an actual experimental realization of the KQR -- trapped
ultracold atoms in the field of a modulated laser standing wave~\cite{Moore}.
The mapping of the KQR to the quasi-random 1d~Anderson model has been
done in Ref.~\cite{Fishman}, a similar analogy was exploited by Gefen and
Thouless~\cite{Thouless} to demonstrate the DL in a mesoscopic disordered
ring threaded by a magnetic flux growing linearly in time.
In Ref.~\cite{Casati90} an analogy between the KQR and band random matrices
was pointed out, the latter have been reduced to a 1d~nonlinear
$\sigma$~model~\cite{Mirlin}. In Ref.~\cite{Altland96} the direct
correspondence between the KQR and a 1d~nonlinear $\sigma$~model was
demonstrated.

On the other hand, numerical simulations for a
$\delta$-kicked particle in an infinite potential well that differs from
the KQR only by the boundary conditions, revealed no DL~\cite{chinese}.
This example shows that DL is not a consequence of the one-dimensional
character of the energy space but it depends on the details of both the
unperturbed system and the perturbation. It is clear that the results on
DL~\cite{Fishman} obtained in the framework of the orthodox
kicked rotor (standard map) model~\cite{Casati79} cannot be automatically
applied to a generic quantum mechanical system under arbitrary
time-dependent perturbation. Even less clear is the status of a peculiar
KQR model with the time-dependent perturbation characterized by three
incommensurate frequencies where the phenomenon similar to the
3d Anderson localization-delocalization transition has been
found numerically~\cite{Shep-inc}.

The most general assumption about a complex quantum-mechanical system
would be the randomness of
the Hamiltonian. It is well-known~\cite{Efetov} that at energies smaller than the
Thouless energy~$E_{c}$ description of complex quantum systems falls into
one of the three universality classes, each corresponding to a Gaussian
ensemble of random matrices. Considering a non-magnetic electron
system with the spin-rotation symmetry,
we arrive at Eq.~(\ref{Ht=}) with $\hat{H}_{0}$
and $\hat{V}$ from the Gaussian orthogonal ensemble (GOE).
An analogous Hamiltonian was considered in connection with laser pumping of
complex organic molecules~\cite{Akulin77}, but DL was not
a target at that stage.
The problem of DL in systems described by the Hamiltonian~(\ref{Ht=})
has been addressed in Ref.~\cite{Wilkinson90}. However, most of results
of this study are qualitative in character and DL has been demonstrated
numerically only for the case where the typical amplitude of harmonic
perturbation $V$ and its frequency $\omega$ are of the order of the mean
level spacing $\delta$ for the unperturbed system.
This numerics has also shown that DL
is unstable with respect to adding even a small
amount of noise in the time dependence.

In this Letter we develop an {\it analytical}\/ theory of DL for a closed
system described by the Hamiltonian~(\ref{Ht=}) with $\hat{H}_{0}$ and
$\hat{V}$ from the GOE of random matrix theory (RMT) and a general
time dependence $\phi(t)$.
We show that in this model the existence of DL and even
the classical ohmic Joule regime with constant absorption rate
depends strongly on the character of $\phi(t)$.
For periodic $\phi(t)=\sum_{n}A_{n}e^{-in\omega t}$
with $|A_{n}|^{2}$ decreasing as $1/n^{3}$ or slower, the absorption
rate diverges and the ohmic
regime does not exist in the limit of infinite matrix size $N$.
In particular, this happens if $\phi(t)$ is a periodic
$\delta$-function as in the orthodox KQR model.
This result alone demonstrates the difference between the
KQR and the time-dependent random matrix theory.

For a {\em periodic}\/ $\phi(t)$ with rapidly decreasing $A_{n}$,
we obtain that the ohmic Joule absorption rate~$W_0$
is modified by the quantum interference effects responsible for the DL.
We calculate the first (one-loop) quantum correction,
which determines the {\it weak}\/ DL,
and show that it is only nonzero if the condition
\begin{equation}
\label{T-rev}
  \phi(-t+\tau)=\phi(t+\tau)
\end{equation}
is fulfilled for a certain choice of $\tau$.
In this case the correction $\delta W(t)\propto \sqrt{t}$ grows
with the time $t$ of the action of the time-dependent perturbation,
like in the quasi-1d Anderson localization in the {\it orthogonal}
symmetry class.
At a certain time scale $t_*$ the first correction $\delta W(t_*)$
becomes comparable with the Joule absorption rate $W_0$ indicating
a crossover from the weak to strong localization in the energy space.
On the other hand, if dephasing time $t_\varphi$ is shorter than $t_*$,
then dissipation remains ohmic, with the rate being smaller than $W_0$
by the value of $\delta W(t_{\varphi})$.

The condition (\ref{T-rev}) is a generalization of the
{\it time-reversal} symmetry condition $\phi(t)=\phi(-t)$
for the case when a shift of the time origin does not matter~\cite{Wang}.
For periodic functions with several harmonics Eq.~(\ref{T-rev})
can be fulfilled only for special choice of relative phases.
However {\it any}\/ pure harmonic function obeys this
condition and thus the monochromatic perturbation is rather an {\it exception}
than a paradigm of a periodic perturbation.
For generic periodic functions which do
not obey Eq.~(\ref{T-rev}) the quantum correction
emerges in the two-loop approximation,
as in systems of the {\em unitary}\/ symmetry class.

Finally, we find that
for $\phi(t)$ being a sum of $d$~harmonic functions with
{\it incommensurate} frequencies the first quantum correction
to the absorption rate is
similar to the weak localization correction to conductivity of a
$d$-dimensional
disordered system of the {\it orthogonal} symmetry class. This gives an
analytical support
of the numerical observation of Ref.~\cite{Shep-inc} for the 
KQR model.

\begin{figure}
\begin{center}
\parbox{7.22cm}{\psfig{figure=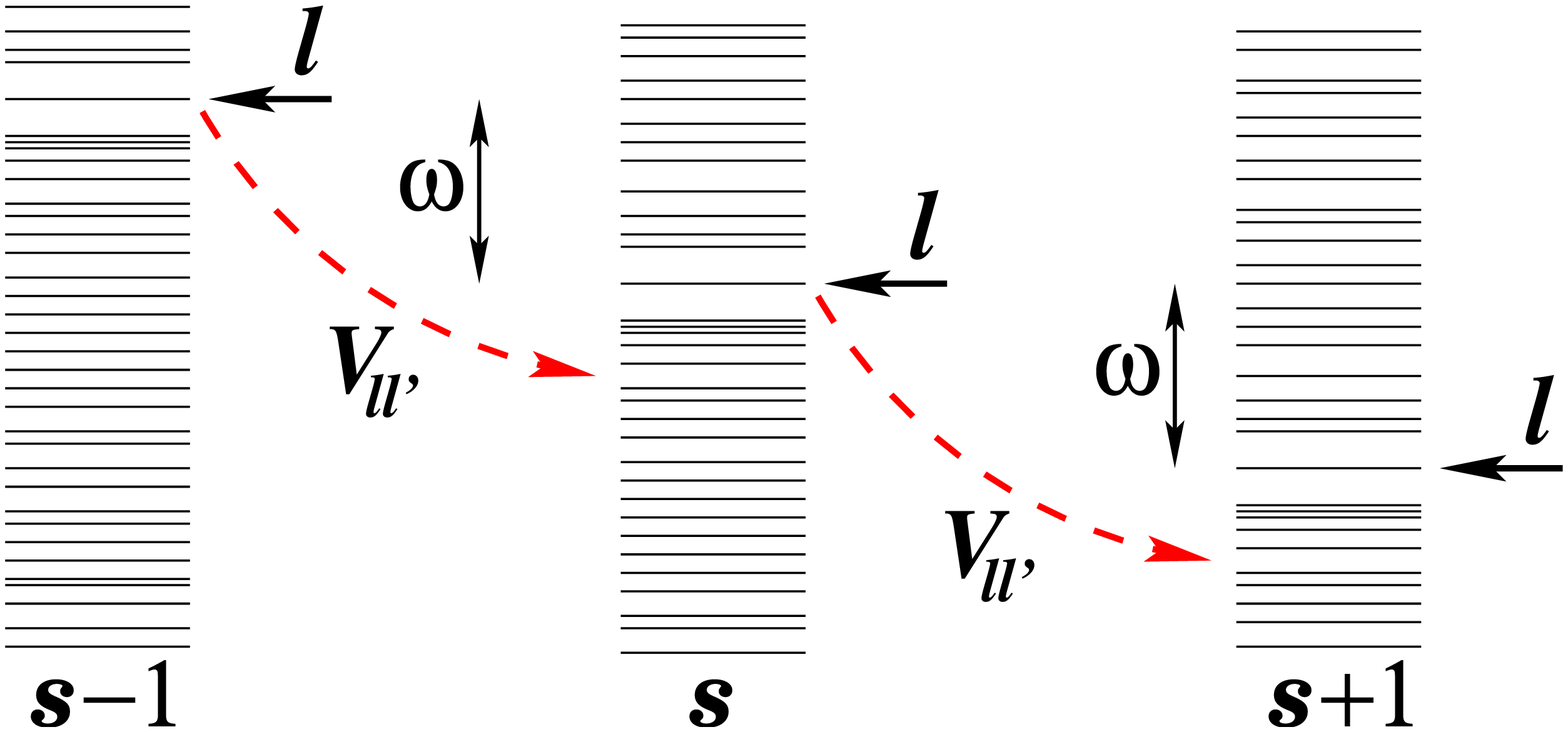,width=7.22cm,height=3.34cm}}
\end{center}
\caption{
Lattice analogy for a monochromatic perturbation.}
\label{F:lattice}
\end{figure}

{\em Qualitative picture and mapping.}---%
The similarity and difference between DL and AL,
as well as between KQR and RMT can be qualitatively
understood using the exact correspondence between a
quantum system under a (multi-)perio\-dic time-dependent perturbation
and a tight-binding lattice model with the time-independent
Hamiltonian. Consider a system with energy
levels (``orbitals'')~$E_l$ under a harmonic perturbation
$V_{ll'}e^{-i\omega{t}}+V_{l'l}^*e^{i\omega{t}}$.
As follows directly from the Schr\"odinger equation, its time evolution
can be expressed in terms of eigenfunctions and eigenvalues
of another system (see Fig.~\ref{F:lattice}), obtained from the original one by
replicating it into a one-dimensional lattice and shifting the
levels of each consequent site by~$\hbar\omega$ so that the energy
of the ``orbital''~$l$ on the site~$s$ is given by $E_{l,s}=E_l-s\hbar\omega$,
and introducing the coupling between the $l$th orbital of the site~$s$ and the
$l'$th orbital of the site~$s+1$ by the {\em stationary} perturbation matrix
element~$V_{ll'}$. In
the same way one can show that higher harmonics in the perturbation
($2\omega,3\omega\,\ldots$) correspond to the coupling to next
neighboring sites (second, third, {\it etc.}), while the presence of several
incommensurate frequencies $\omega_1,\ldots,\omega_d$ requires a
$d$-dimensional lattice with sites $\vec{s}=(s_1,\ldots\,s_d)$,
on-site energies $E_l-s_1\omega_1-\ldots-s_d\omega_d$, and the matrix
element~$V_{ll'}^{(i)}$ at the $i$th frequency corresponding to the
coupling along the $i$th dimension.

The $d$-dimensional space with ``orbitals'' on different
``sites'' coupled by the matrix elements of perturbation, arises quite
naturally from the notion of {\it quasi-energy}~\cite{quasienergy}.
However, this model is far from the conventional Anderson model,
where there is only one orbital per site with a random energy and from
the Wegner's $N$-orbital model \cite{Wegner} where there is a fixed number
of orbitals with random energies and random hopping integrals.
The point is that being a copy
of the original spectrum, the set of orbitals' energies are not independent at
different sites. Another difference is that the number of orbitals is infinite
in the relevant limit of the infinite matrix size.

From the above mapping it is clear why slowly decreasing amplitudes of harmonics
kill localization in RMT. It happens because of the possibility of
long-range hops over the
sites. It is also clear why the long-range hops are not so dangerous for
localization
in the KQR case. The point is that in the orthodox KQR model,
the time-dependent
perturbation is coupled to $\cos\theta$ which in the basis of eigenfunctions
$e^{i l\theta}$ of the unperturbed system corresponds to matrix
elements $V_{l,l\pm 1}$
that may connect only neighboring orbitals. For long enough distance between
sites the coupled orbitals run out of resonance and the hopping is suppressed.
Should the basis  change to
$\sin( l \theta)$ as in the case of infinite potential well in
Ref.~\cite{chinese},
the matrix elements $V_{l,l'}\propto 1/|l-l'|$ become long-range in
the orbitals' space,
so that the resonance hopping between remote sites may still occur,
though only between remote orbitals. For the RMT case $V_{l,l'}$ does not
decrease with the distance $|l-l'|$ and the condition
for the long-range resonance hopping is further improved.

This consideration shows that the KQR with $\delta$-kicks and the
time-dependent RMT with harmonic $\phi(t)$ are in fact two {\it
complementary} models. In the former {\it remote sites} are connected
but only for {\it neighboring orbitals},
while in the latter {\it remote orbitals} are connected only on
{\it neighboring sites}.
Sometimes -- as in the case of several harmonics with incommensurate
frequencies -- this complementarity makes two  models
equivalent but this equivalence is fragile and needs to be
checked in every particular case.

{\em Description of the formalism.}---Consider a closed system of
non-interacting fermions with the single-par\-ticle Hamiltonian~(\ref{Ht=}),
$\hat{H}_0$ and $\hat{V}$ being real symmetric random $N\times{N}$
matrices distributed with probabilities
$$
  \mathcal{P}_{H_0} \propto
    \exp\left[-\frac{\pi^2\tr{H}_0^2}{4N\delta^2}\right],
\qquad
  \mathcal{P}_{V} \propto
    \exp\left[-\frac{\pi\tr{V}^2}{4\Gamma\delta}\right],
$$
respectively, $\delta$~being the mean level spacing at the center of the band,
and $\Gamma$~measuring the sensitivity of the energy levels $E_l$ to the
variation of $\phi$:
$\langle (\partial E_l/\partial\phi)^2\rangle = (2/\pi)\Gamma\delta$.
Such a form of time-dependent random matrix
theory corresponds to a quantum dot under a perturbation with characteristic
frequencies $\omega\ll{E}_c$~\cite{Kanzieper}.

To treat the dynamical problem specified by the Hamiltonian (\ref{Ht=}),
we employ the nonlinear $\sigma$-model approach recently developed
in Ref.~\cite{Misha} on the basis of the Keldysh nonequilibrium
formalism~\cite{Horbach,Andreev}.
In the limit $N\to\infty$, the effective action
of the $\sigma$-model reads:
\begin{eqnarray}
\nonumber
 S[Q] \! &=& \!\! \int\left[\frac{\pi i}{2\delta}
  \tr\{i\tau_3\sigma_0\delta_{tt'}\partial_tQ_{tt'}\}
  \right.
\\
\label{sigmaction=}
&& \!\! +\left.\frac{\Gamma}{8\delta}\,[\phi(t)-\phi(t')]^2\,
  \tr\{Q_{tt'}Q_{t't}\}\right]
dt\,dt'.
\end{eqnarray}
Here $Q_{tt'}$ is a $4\times{4}$ matrix in the direct product of the
$2\times{2}$ particle-hole and Keldysh spaces.
Pauli matrices in these spaces are denoted by $\tau_i$~and~$\sigma_i$,
respectively.
The first term in $S$ is the standard random-matrix action~\cite{Altland}
responsible for the whole spectral statistics, while the second -- kinetic
-- term accounts for the effects of the time-dependent perturbation.
The matrix~$Q$ is subject to the constraints
$(Q^2)_{tt'}=\tau_0\sigma_0\delta_{tt'}$ and
$\tau_2\sigma_1Q\sigma_1\tau_2=Q^T$,
where the product and the transpose involve the time arguments too.
The saddle point of the action~(\ref{sigmaction=}) is
\begin{equation}
  \Lambda_{tt'} =
    \left(\begin{array}{cc}
      \delta_{tt'} & 2F_{tt'}^{(0)} \\ 0 & -\delta_{tt'}
    \end{array}\right) \otimes \tau_3,
\label{Lambda}
\end{equation}
where the function $F_{tt'}^{(0)}$ satisfies the kinetic equation
\begin{equation}
\label{saddleq=}
  \left[\partial_t+\partial_{t'}+\Gamma\,(\phi(t)-\phi(t'))^2\right]
  F_{tt'}^{(0)}=0\:.
\end{equation}
$F_{tt'}$ is the electron distribution function in the
time representation; a more familiar quantity is its Wigner transform
$F_E(t) = \int d\tau e^{iE\tau} F_{t+\tau/2,t-\tau/2}$.
In equilibrium with temperature~$T$ it is $F_E=\tanh(E/2T)$.
Out of equilibrium it satisfies the Wigner-transformed Eq.~(\ref{saddleq=}),
which after averaging over fast oscillations in $t$ reduces
to the diffusion equation in the energy space:
\begin{equation}
\label{diffusion=}
  \left[\partial_t-D\,\partial_{E}^2\right] F^{(0)}_E(t) = 0,
\qquad
  D=\Gamma\,\overline{(d\phi/dt)^2}\:,
\end{equation}
the overline meaning the average over time. Equation~(\ref{diffusion=})
gives ohmic Joule absorption rate
\begin{equation}
\label{W0}
  W_0 = D/\delta .
\end{equation}
The saddle-point expression (\ref{W0}) is valid provided that
(i)~the perturbation is sufficiently fast,
$\partial_t\phi\gg \delta^{3/2}\Gamma^{-1/2}$,
which is the anti-adiabaticity condition
ensuring that the spectrum is quasi-continuous
and (ii) interference effects responsible for DL are neglected.

The perturbative correction to Eq.~(\ref{W0}) for the case
of linear bias $\phi(t)=vt$ was calculated in Ref.~\cite{Misha}.
There, the interference is ineffective and quantum corrections
describe the high-$v$ tail of the crossover from the Kubo
to the Landau-Zener regimes of dissipation~\cite{Wilkinson88}.
Contrary, in this Letter we concentrate on the case of large
$\partial_t\phi\gg \delta^{3/2}\Gamma^{-1/2}$ but bounded $\phi(t)$,
when the saddle-point approximation is invalidated by
interference effects.

Quantum corrections to the mean-field absorption rate~(\ref{W0})
can be obtained in the regular way by expanding
over Gaussian fluctuations near the saddle point.
The deviation of the $Q$~matrix from $\Lambda$ is
parametrized~\cite{Misha} by the diffuson and cooperon modes
$b_{tt'}$ and $a_{tt'}$ with the bare propagators:
$\langle b_{t_+t_-}b^*_{t_+'t_-'}\rangle =
(2\delta/\pi)\,\delta(\eta-\eta')\,\cD_{\eta}(t,t')$
and $\langle a_{t_+t_-}a^*_{t_+'t_-'}\rangle =
(\delta/\pi)\,\delta(t-t')\,\cC_t(\eta,\eta')$,
where
\begin{eqnarray}
  &&\cD_{\eta}(t,t')=\theta(t-t')\,\exp\left[-\int_{t'}^t
  \gamma_{\eta}(\tau)\,d\tau\right],
\\
  &&\cC_t(\eta,\eta')=\theta(\eta-\eta')\,\exp\left[-\frac{1}{2}
  \int_{\eta'}^{\eta}\gamma_{\xi}(t)\,d\xi\right],
\end{eqnarray}
and we have denoted $t_{\pm}=t\pm\eta/2$, $t_{\pm}'=t'\pm\eta'/2$,
and $\gamma_{\eta}(t)\equiv\Gamma\,[\phi(t_+)-\phi(t_-)]^2$.

In the presence of fluctuations, the average matrix $\left<Q\right>$
still has the form (\ref{Lambda}) but with the saddle-point $F^{(0)}$
substituted by the renormalized electron distribution $F$
which determines the energy absorption rate:
\begin{equation}
  W(t) \equiv \partial_t \langle{E(t)}\rangle =
  - \frac{i\pi}{\delta} \lim_{\eta\to0}
    \partial_{t}\partial_{\eta} F_{t+\eta/2,\,t-\eta/2}\:.
\end{equation}

The one-loop quantum correction to $\langle{Q}\rangle$ contains
a diffuson and one cooperon loop coupled by a Hikami box.
We evaluate this diagram for a generic perturbation $\phi(t)$
switched on at $t=0$.
Using the asymptotics $F_{t_+t_-}\sim(i\pi\eta)^{-1}$
following from the property $\lim_{E\rightarrow\pm\infty}F_E(t)=\pm{1}$,
we obtain for the absorption rate:
\begin{equation}
\label{dissiprate=}
  W(t) = W_0 + \frac{\Gamma}{\pi}\int_0^t
    \phi'(t)\,\phi'(t-\xi)\,\cC_{t-\xi/2}(\xi,-\xi)\,d\xi .
\end{equation}
The second term in Eq.~(\ref{dissiprate=}) is the one-loop quantum
interference correction to the ohmic Joule heating (\ref{W0}).
Expression~(\ref{dissiprate=}) is the main result of this part of the
paper and will be the base for the subsequent considerations.
It can be also obtained from the conventional diagrammatic
technique~\cite{Vavilov,Kanzieper}.

{\em Results.}---%
First, we consider a {\em periodic}\/ perturbation:
$\phi(t)=\sum_{n}A_n\cos(n\omega{t}-\varphi_n)$,
with the diffusion coefficient $D=(\omega^2\Gamma/2)\sum_nn^2A_n^2$.
If $A_n$~decrease as~$n^{-3/2}$ or slower (e.~g. for $\delta$-kicks),
the diffusion coefficient and dissipation rate diverge.
This divergency is cut either by a finite size $N$ of the matrix
or by the interaction effects that is beyond our analysis.

\begin{figure}
\begin{center}
\parbox{7.68cm}{\psfig{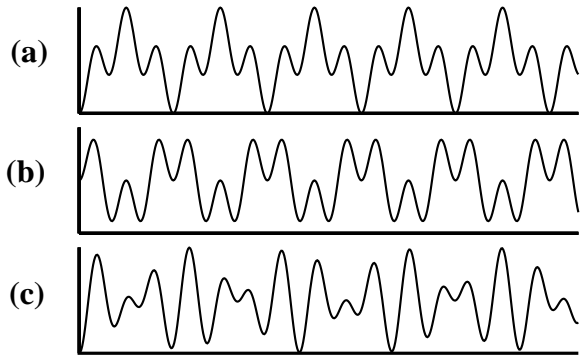}}
\end{center}
\caption{\label{nodeph:}
Perturbation-induced dephasing rate [the coefficient at~$\xi$ in the argument
of the exponential in Eq.~(\protect\ref{cooperontxi=})] vs.\ $\xi$:
(a)~periodic $\phi(t)$ obeying Eq.~(\ref{T-rev}): a regular array of
zeros
;
(b)~general periodic $\phi(t)$: a gap%
;
(c)~quasi-periodic $\phi(t)$ with two incommensurate frequencies:
neither a gap nor a regular array of zeros%
.
}
\end{figure}

To study the long-time, period-averaged dynamics at
$t,\xi\gg 1/\omega$ we can approximate
\begin{equation}
\label{cooperontxi=}
  \cC_{t-\xi/2}(\xi,-\xi) \approx
  e^{-2\Gamma\xi\sum_{n} A_n^2\sin^2[n\omega(t-\xi/2)-\varphi_n]} .
\end{equation}
For a particular choice of phases, $\varphi_n=n\varphi$, there exists
a set of points $\xi_k=2t-2(\varphi+\pi{k})/\omega$ with integer~$k$ where
the cooperon (\ref{cooperontxi=}) is equal to unity (Fig.~\ref{nodeph:}a).
Existence of such {\em no-dephasing points}~\cite{Wang} is equivalent
to the generalized time-reversal symmetry (\ref{T-rev}) of the perturbation.
At large~$\xi$ only $\xi\approx\xi_k$ contribute to the
integral (\ref{dissiprate=}) (otherwise the cooperon is exponentially small)
which then can be calculated with the steepest descent method.
Performing summation over the no-dephasing points $\xi_k$ we obtain
a negative and growing in time quantum interference correction
to the ohmic absorption rate (\ref{W0}):
\begin{equation}
\label{dotEper=}
  \frac{W(t)}{W_0} = 1 - \sqrt{\frac{t}{t_*}} ,
\qquad
  t_* = \frac{\pi^3\Gamma\overline{n^2}}{2\delta^2}
\end{equation}
where $\overline{n^2}\equiv\sum_nn^2A_n^2$
and the limit $t\gg{1}/\omega,{1}/\Gamma$ is implied.
The $\sqrt{t}$ dependence is remarkably similar to the
$\sqrt{t_{\varphi}}$~dependence of the first quantum correction to the
conductivity of a particle in a quasi-one-dimensional disordered
sample with the phase relaxation time~$t_{\varphi}$~\cite{GLH}.
In our case the relative quantum correction
becomes comparable to unity at time $t_*$.
This is an indication of DL at the characteristic energy scale
$E_*\sim\sqrt{Dt_*}\sim\omega\Gamma\overline{n^2}/\delta$.
Taking
$\delta=3$~$\mu$eV, $\hbar\omega=40$~$\mu$eV ($\omega/2\pi\approx{10}$~GHz),
$\Gamma\overline{n^2}=10$~$\mu$eV (corresponding to the microwave electric
field 
of a few~V/m over the dot size of $1$~$\mu$m)
we obtain $t_*\sim{10}$~ns, $E_*\sim{400}$~$\mu$eV~$\sim{5}$~K.

When the generalized time-reversal symmetry (\ref{T-rev}) is absent
for~$\phi(t)$, the dephasing rate
is positive and separated from zero by a finite gap (Fig.~\ref{nodeph:}b).
In this case the integral over~$\xi$ converges exponentially, so the first
quantum correction stays small even at $t\rightarrow\infty$, as in the
systems of the unitary symmetry class.

Next, we consider the case of $d$~{\em incommensurate}\/ frequencies
$\omega_n$: $\phi(t)=\sum_{n=1}^d A_n\cos(\omega_nt-\varphi_n)$,
for which $D=(\Gamma/2)\sum_{n=1}^d\omega_n^2A_n^2$.
For incommensurate frequencies the relationship between the
phases~$\varphi_n$ does not matter.
The reason is that in this case there can be only one no-dephasing point
at most which cannot make a decisive contribution to the integral, in
contrast to the periodic case with an infinite number of such points.
On the other hand, in the incommensurate case the dephasing rate can become
arbitrarily small (pseudo-gap) in an infinite number of points
regardless of phases (Fig.~\ref{nodeph:}c),
and the integral is dominated by these points.

To calculate the first quantum correction in Eq.~(\ref{dissiprate=})
we use the cooperon (\ref{cooperontxi=}) with $n\omega$ substituted
by $\omega_n$, expand it into a $d$-dimensional sum involving the
modified Bessel functions $I_\nu(z)$, and average over $t$
which is significantly simplified due to incommensurability.
The analytical expression for the case of arbitrary relation
between the amplitudes $A_n$ is bulky.
We write the resulting expression valid at $t\gg{1}/\omega,{1}/\Gamma$
for the case when all $A_n=1$:
\begin{equation}
\label{multiharm=}
  \frac{W(t)}{W_0}=1-\frac{\delta}{\pi\Gamma}
  \int_{0}^{\Gamma{t}}dz\,
  e^{-z{d}}\,[I_0(z)]^{d-1}\,\frac{dI_0(z)}{dz} .
\end{equation}
Using the asymptotic form
$I_0(z)\approx dI_0(z)/dz\approx e^z/\sqrt{2\pi{z}}$ at $z\gg{1}$,
we recover Eq.~(\ref{dotEper=}) for a monochromatic perturbation.
For $d=2$, the relative correction is $(\delta/2\pi^2\Gamma)\ln\Gamma t$,
whereas for $d>2$ it saturates $\propto t^{1-d/2}$ in the limit of large $t$,
in complete analogy with the behavior of the quantum correction
of Ref.~\cite{GLH} in $d$~dimensions.

In conclusion, we have developed an analytical approach based on the {\it
zero-dimensional, time-dependent} nonlinear sigma-model and obtained the
{\it weak dynamical localization} in complex quantum systems under
time-dependent
perturbation described by the {\it random matrix theory}. The character of
energy absorption in such systems
is determined entirely by the frequency spectrum of a time-dependent
perturbation. In particular we obtained no DL for the time-periodic
$\delta$-function perturbation, and the dynamical localization corections 
similar to the $d$-dimensional weak localization corrections to
conductivity if
the perturbation is a sum of $d$~incommensurate harmonic functions.

The authors are grateful to V.~I.~Fal'ko, A.~D.~Mirlin, and
O.~M.~Yevtushenko for helpful discussions, and to M.~G.~Vavilov for
providing a copy of his Ph.D. thesis.
M.~A.~S. acknowledges financial support from the SCOPES program of
Switzerland, the Dutch Organization for Fundamental Research (NWO),
the Russian Foundation for Basic Research under grants 01-02-17759
and 02-0206238, the program "Quantum Macrophysics" of the Russian Academy
of Sciences, the Russian Ministry of Science, the Russian Science Support
Foundation, and thanks Abdus Salam International Centre for Theoretical
Physics for hospitality and support.

\end{document}